\documentclass[%
 reprint,
showpacs,
 amsmath,amssymb,
 aps,prl,
]{revtex4-1}

\usepackage{graphicx}% Include figure files
\usepackage{dcolumn}% Align table columns on decimal point
\usepackage{bm}% bold math
\usepackage{color}
\begin{document}

\preprint{APS/123-QED}

%\title{A topologically-based state description for equilibrium and dynamic \\
%capillary pressure in two-fluid porous media}% Force line breaks with \\
\title{A geometric state function for two-fluid flow in porous media}% Force line breaks with \\

\author{James E. McClure} 
\affiliation{Virginia Polytechnic Institute \& State University, Blacksburg} 
\author{Ryan T. Armstrong}
\affiliation{University of New South Wales, Sydney}
\author{Mark A. Berrill} \thanks{This manuscript has been authored by UT-Battelle, LLC under Contract No. DE-AC05-00OR22725 with the U.S. Department of Energy. The United States Government retains and the publisher, by accepting the article for publication, acknowledges that the United States Government retains a non-exclusive, paid-up, irrevocable, world-wide license to publish or reproduce the published form of this manuscript, or allow others to do so, for United States Government purposes.  The Department of Energy will provide public access to these results of federally sponsored research in accordance with the DOE Public Access Plan (http://energy.gov/downloads/doe-public-access-plan).}
\affiliation{Oak Ridge National Laboratory, Oak Ridge}
\author{Steffen Schl\"uter}
\affiliation{Helmholtz-Centre for Environmental Research - UFZ, Halle (Saale)}
\author{Steffen Berg}
\affiliation{Shell Global Solutions International B.V.
Grasweg 31,
1031HW Amsterdam,
The Netherlands}
\author{William G. Gray}
\author{Cass T. Miller}
\affiliation{University of North Carolina at Chapel Hill, Chapel Hill}

\date{\today}% It is always \today, today,

\begin{abstract}
Models that describe two-fluid flow in porous media suffer from a widely-recognized problem that the constitutive relationships used to predict capillary pressure as a function of the fluid saturation are non-unique, thus requiring a hysteretic description. As an alternative to the traditional perspective, we consider a geometrical description of the capillary pressure, which relates the average mean curvature, the fluid saturation, the interfacial area between fluids, and the Euler characteristic. The state equation is formulated using notions from algebraic topology and cast in terms of measures of the macroscale state. Synchrotron-based X-ray micro-computed tomography ($\mu$CT) and high-resolution pore-scale simulation is applied to examine the uniqueness of the proposed relationship for six different porous media. We show that the geometric state function is able to characterize the microscopic fluid configurations that result from a wide range of simulated flow conditions in an averaged sense. The geometric state function can serve as a closure relationship within macroscale models to effectively remove hysteretic behavior attributed to the arrangement of fluids within a porous medium. This provides a critical missing component needed to enable a new generation of higher fidelity models to describe two-fluid flow in porous media.

\end{abstract}

\maketitle

%\tableofcontents
%\section{\label{sec:intro}Introduction}

Two-phase extensions of Darcy's law were introduced on a phenomenological basis more than a half-century ago \cite{Leverett_41,vanGenuchten_80,Brooks_Corey_66,Mualem_76}. These models
are used routinely to predict the behavior of hydrologic systems, evaluate geologic carbon sequestration, and guide the recovery of oil and gas \cite{juanes2006impact,szulczewski2012lifetime,berg2010miscible}.  
A widely-recognized deficiency of these models is reliance upon non-unique and history-dependent closure relationships. A prominent example is the relation used to describe the capillary pressure as a function of the fluid saturation history \cite{Hassanizadeh_Gray_93,Bear_Rubenstein_etal_11,Amaziane_Milisic_etal_12,Killough_76}. 
It has been hypothesized that history-dependence can be removed from the capillary pressure relation
by incorporating additional state variables
\cite{Hassanizadeh_Gray_93}. Despite experimental and computational efforts over the last two decades, a sufficient set of state variable has not been established to confirm this hypothesis and no general unique state equation exists to desribe capillary pressure.  As a result, traditional, empirical, hysteretic approximations for capillary pressure are still overwhelmingly used \cite{vanGenuchten_80}.

% PARAGRAPH 2 -- existing work on hysteresis
While models must be formulated and applied at macroscopic length scales that range from meters to kilometers, it is known that the microscopic arrangement of fluids within geologic materials has important consequences for fluid flow \cite{Ferer_Bromhal_etal_05,Reeves_Rothman_12,Rozman_Utz_02,Hilfer_Armstrong_etal_15,Tallakstad_Knudsen_etal_09,Hatiboglu_Babadagli_08,Cueto-Felgueroso_Juanes_08,Holtzman_Segre_15,Chevalier_Salin_etal_15,Zhao_etal_2016,Xu_Louge_15}. In particular, the snap-off and entrapment of fluid sub-regions at the pore-scale has been established as an important source of hysteresis \cite{Chatzis_Morrow_83,Held_Celia_01c,joekar2013trapping,Schlueter_Berg_etal_15,Schlueter2016b}. 
The impact of fluid connectivity on the system behavior has led to the development of approaches
that explicitly include the portion of disconnected fluids \cite{Land_68,Lenhard_Parker_87,Hilfer_06,Spiteri_etal_08,Sakai_15}.
Relationships that include additional state
variables, such as interfacial area \cite{Held_Celia_01c,Cheng_Pyrak-Nolte_etal_04,Porter_Wildenschild_etal_10,Joekar-Niasar_Hassanizadeh_11,Peng_Brusseau_12,karadimitriou2014micromodel,joekar2010network}
and more recently the Euler characteristic \cite{Herring_Harper_etal_13,Liu_Herring_etal_17,McClure_Berrill_etal_16b,Armstrong_McClure_etal_16} have also been considered. 

%Paragraph 4
% Clearly state the objectives and purpose of the manuscript
The overall goal of this work is to develop and validate a hysteretic-free geometric state equation 
to describe the macroscale capillary pressure based on invariant measures. The specific objectives of this work are:
(1) to provide a theoretical bases for a geometric state function; (2) to produce a general functional form of the state equation in terms of macroscale measures of the state of a porous medium; 
(3) to define a specific state equation consistent with the general form; (4) to evaluate the state for a wide range of media and flow conditions; (5) to analyze the state data in light of traditional, alternative, and new state equations; and (6) to investigate whether the proposed state equation can describe not only equilibrium but also dynamic states of the system. 

\section{Capillary Pressure}

Mercury porosimetry techniques were developed to probe the geometry of a porous medium by 
modifying the pressure difference between two fluids, $p_c=p_n - p_w$ \cite{Giesche_06}. 
At equilibrium and in the absence of external forces, the pressure forces are balanced 
by capillary forces at points on the interface between fluids, as given in the 
Laplace equation:
\begin{equation}
p_n - p_w = \gamma_{wn} \Bigg(\frac {1}{R_1} + \frac {1}{R_2} \Bigg) \;,
\label{eq:pc}
\end{equation}
where $R_1$ and $R_2$ are the principal radii of curvature determined at points on the surface
and $\gamma_{wn}$ is the interfacial tension between fluids.
Typical porous media have a range of pore sizes, and at the macroscale as the capillary pressure $p^c$ increases the
saturation of the wetting fluid, $s^w$, decreases \cite{Adamson_Gast_97,Pinder_Gray_08}. This reasoning provides the basis for Leverett's J-function and van Genuchten's relation to predict the water retention curve in porous materials \cite{Leverett_41,vanGenuchten_80}. However, the relation $p^c(s^w)$ is material-specific and depends on the system history.
A typical example of a hysteretic capillary pressure relationship is shown in Fig.
\ref{fig:hysteresis}. Different capillary pressures
are observed for a given saturation depending on the flow history, meaning that a functional relationship
$p^c(s^w)$ does not exist. During drainage, higher capillary pressures are needed to force the menisci through
the pore-throats, which are the narrowest parts of the porespace. Capillary pressures along the imbibition curve
are determined by the pore body sizes, which are associated with larger radii of curvature. 
Snap-off and coalescence events occur frequently during displacement and alter the connectivity of the fluids.
As a result, fluid connectivity is recognized as an important component of the system state. Key features of the
capillary pressure relationship correspond with the limits where fluid connectivity breaks down. The residual non-wetting phase saturation corresponds to a state where the non-wetting fluid is completely disconnected and ceases to transmit a pressure gradient across a system. Similarly, the irreducible wetting phase saturation is reached when wetting phase connectivity
breaks down. These values are often used as parameters in empirical functional forms 
used to fit experimental data for capillary pressure and relative permeability to account for the role of fluid connectivity \cite{vanGenuchten_80,Brooks_Corey_66,Mualem_76}. 

\begin{figure}[ht]
\centering
\includegraphics[width=1.0\linewidth]{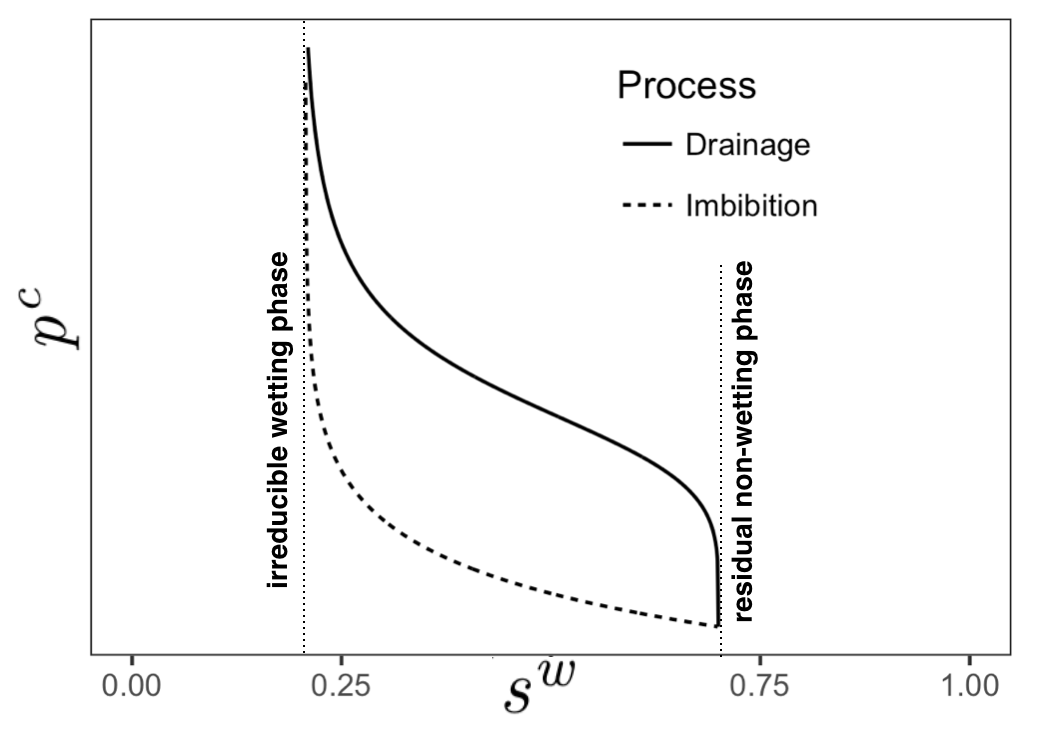}
\caption{The shape of drainage and imbibition curves are determined in part 
by the fluid connectivity. The irreducible wetting phase and residual non-wetting 
phase saturations correspond to limits where fluid connectivity breaks down.}
 \label{fig:hysteresis}
\end{figure}

\section{\label{sec:topo} Geometric State Measures}

% P2 -- average measures of the geometry
Invariant geometric quantities provide a natural mechanism to characterize the complex arrangements of fluid and
solid phases within porous media \cite{Mecke_98,Arns_Knackstedt_etal_04,Hilfer_02,Scholz_12,Ohser_11}. 
The mathematical underpinnings for 
this approach are provided by Hadwiger's theorem, which demonstrates that only four averaged measures,
the Minkowski functionals (MFs),
are needed to characterize the morphology of a three-dimensional object
\cite{Hadwiger1957,Klain_95,Nagel_2000,Serra_1983}. We consider the non-wetting phase domain, $\Omega_n \in \Omega$, 
with boundary $\Gamma_n$, where $\Omega$ is the domain occupied by a porous medium. The Minkowski functionals can be considered
as macroscale invariants that are obtained directly from microscale invariants using integral geometry.
The first pair of Minkowski functionals are the volume and the surface area,
\begin{eqnarray}
M_0^n &=& \lambda(\Omega_n)  = \int_{\Omega_n} dr 
%=\epsilon^{{n}} V,
\label{eq:M0} \\
M_1^n &=&  \lambda(\Gamma_n) = \int_{\Gamma_n} dr\;, 
% = (\epsilon^{{wn}} + \epsilon^{{ns}})V,
\label{eq:M1} 
\end{eqnarray}
where $\lambda$ denotes the Lesbesgue measure.
The second pair of Minkowski functionals are averages of the two microscale surface invariants on $\Gamma_n$;
these are the mean curvature and the Gaussian curvature. The associated Minkowski functionals are
\begin{eqnarray}
M_2^n &=& 
\int_{\Gamma_n} \Bigg(\frac{1}{R_1} +  \frac{1}{R_2} \Bigg )dr
\quad\mbox{and}
%(J_w^{wn}\epsilon^{{wn}} + J_s^{ns}\epsilon^{{ns}})V,
\label{eq:M2} \\
M_3^n &=& 
\int_{\Gamma_n} \frac{1}{R_1 R_2} dr\;.
%+ \int_{\partial \Gamma_n} \kappa_g dr.
%= \chi^{{n}}
\label{eq:M3} 
\end{eqnarray}
%$M_3^n$ is the total curvature of the boundary surface, which includes a contribution from the integral of the Gaussian curvature of $\Gamma_n$ plus the integral of the geodesic curvature $\kappa_g$ along the the surface boundary curve $\partial \Gamma_n$. For a smooth, closed surface the boundary source term is identically zero. A physically interesting case is the contribution of the contact angle to the total curvature, which will be treated subsequently. 

\begin{figure}[ht]
\centering
\includegraphics[width=1.0\linewidth]{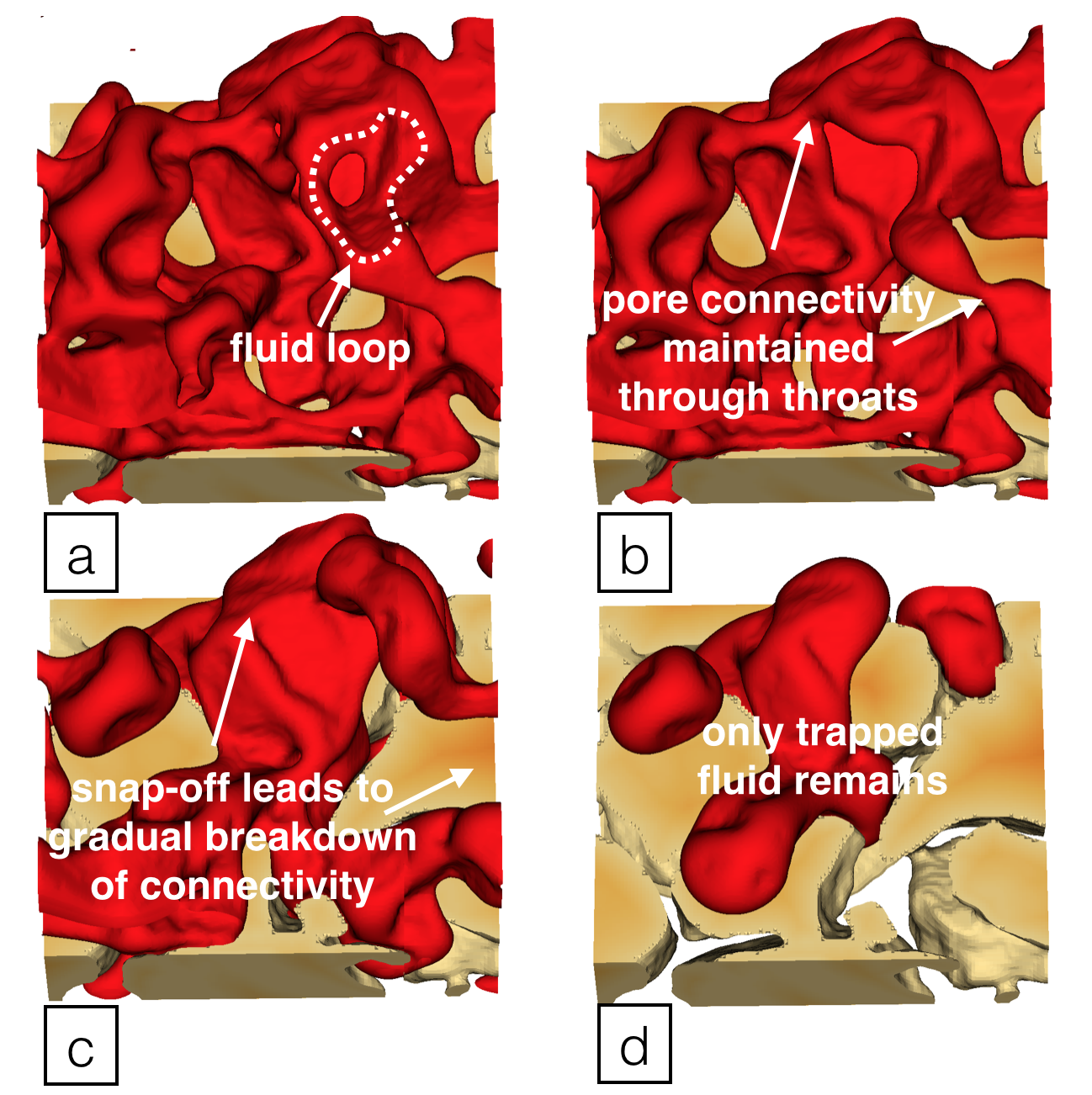}
\caption{Sequence showing the pore-level evolution of the non-wetting fluid geometry during imbibition of the wetting fluid. 
As the volume fraction of the non-wetting fluid volume decreases, corresponding changes in the
surface area, mean curvature and connectivity also occur.
}
 \label{fig:twophase}
\end{figure}

Each of the MFs measure the size of $\Omega_n$ in various dimensions: volume ($\ell^3$), surface area ($\ell^2$), integral mean curvature ($\ell$), and the Euler characteristic 
\begin{equation}
\chi^{{n}} = \frac{M_0^n}{4 \pi}\;. 
\end{equation}
The Euler characteristic relates to the total curvature by the Gauss-Bonnet theorem, and serves as an average measure for the connectivity of the fluid based on the possible channels and disconnected regions \cite{Federer_1959,Hilfer_02}. Topological
theory links the Euler characteristic to the number of
connected components $B_0^n$, loops $B_1^n$ and cavities $B_2^n$,
\begin{equation}
\chi^{{n}} = B_0^n - B_1^n + B_2^n\;. 
\label{eq:topology}
\end{equation}
The quantities $B_0^n$, $B_1^n$ and $B_2^n$ are topological invariants known 
as the Betti numbers. The pore-level view illustrated in Fig. \ref{fig:twophase} provides insight into the relationship
between the Betti numbers and the non-wetting fluid connectivity. 
In this case, the volume fraction of the non-wetting fluid is decreasing from (a)--(d) due to imbibition of the
wetting fluid. The geometric evolution of the non-wetting fluid occurs based on incremental changes 
to the volume, surface area, curvature and connectivity. These changes are coupled.
In the well-connected system shown in Fig. \ref{fig:twophase} (a), redundant connectivity
is evident based on the presence of loops. As the fluid volume fraction decreases,
connections made through the pore throats are broken due to snap-off, 
eventually leaving only non-wetting fluid components that are disconnected from each other. The Euler characteristic 
quantifies these effects.

\begin{figure}[ht]
\centering
\includegraphics[width=0.7\linewidth]{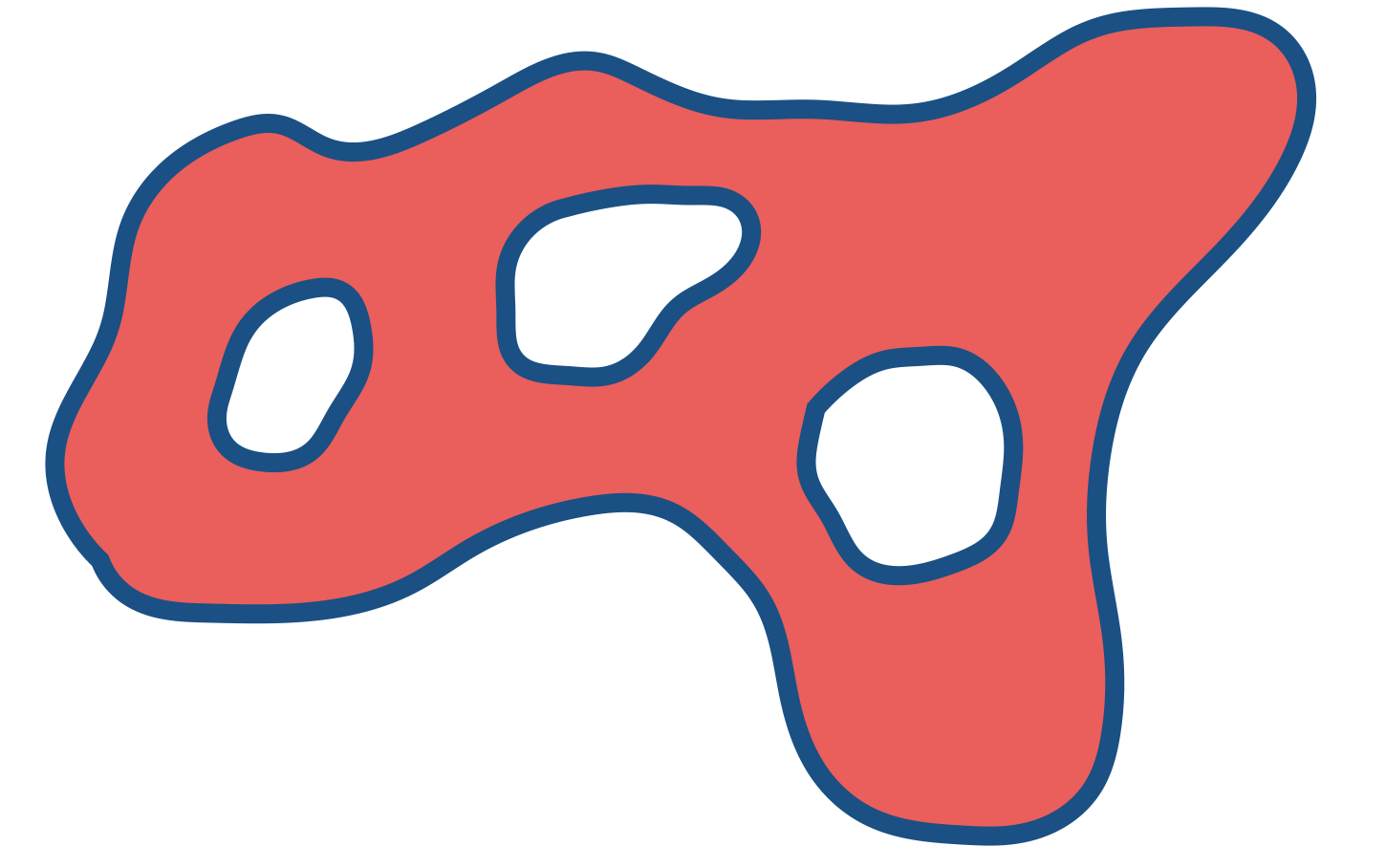}
\caption{Two-dimensional analog for Eq. \ref{eq:steiner-2}: the change in area  (blue region) obtained by rolling a small ball around the boundary for a 2D object (red region) is determined by
the shape of the red object boundary, and can be expressed in terms of averaged measures of the boundary shape. 
}
 \label{fig:parallel-set}
\end{figure}

% Part II - entities and what quantities are independent
The basis for characterization of the non-wetting fluid is provided by a 
generalized form of Steiner's formula that applies for sets with positive reach \cite{Federer_1959}
\begin{equation}
%f(M_0^n,M_1^n,M_2^n,M_3^n) = 0,
\lambda(\Omega_n \oplus \varsigma_{\delta}) - \lambda(\Omega_n) = \sum_{i=1}^3 a_i M_i^n \delta^i\;,
\label{eq:steiner}
\end{equation}
where $\varsigma_{\delta}$ is a spherical ball with radius
$\delta$ and $\oplus$ denotes the Minkowski sum. Steiner's formula predicts
the volume of the parallel set $\Omega_n \oplus \varsigma_{\delta}$ 
for a particular $\Omega_n$. The coefficients $a_i$ are 
determined by the shape of $\Omega_n$. Eq. \ref{eq:steiner} shows that
sufficiently small changes in the volume of $\Omega_n$ 
are locally-smooth and continuous functions of the MFs \cite{Federer_1959,Klain_95}. 
The assumption that $\Omega_n$ has positive reach means that for some positive
$\delta$ the ball $\varsigma_{\delta}$ could be rolled around the boundary without intersecting $\Omega_n$. The non-wetting fluid will in general meet this criterion. 
As an illustrative example, consider the two-dimensional object
shown in Fig. \ref{fig:parallel-set}. In this case, the blue region is produced by rolling
a small ball around the boundary of the red object (i.e. $\Omega_n$). The area
of the blue region is determined only by the properties of the red region boundary and the size of the ball.
This is the meaning of Eq. \ref{eq:steiner}. The red object has positive reach because the
blue region does not overlap with the red object at any point along the boundary. 
However, the red object is not an convex set; if a large enough ball were rolled along the boundary, the holes in the middle of the object would be closed by the operation 
$\Omega_n \oplus \varsigma_\delta$. This would irrevocably alter the topology of the object.
For this reason Eq. \ref{eq:steiner} will only hold if the ball $\varsigma_\delta$ is sufficiently
small. When the boundary of the wetting and solid phases include grain contacts, 
the associated sets will not have positive reach.
This motivates our choice to characterize the non-wetting fluid. 

Since the MFs are extensive properties, it is useful to divide Eq. \ref{eq:steiner} by the total volume,
\begin{equation}
\Delta \epsilon^n =
\frac {\lambda(\Omega_n \oplus \varsigma_\delta) - \lambda(\Omega_n)}{V} =  \sum_{i=1}^3 \frac{a_i M_i^n \delta^i}{V}\;,
\label{eq:steiner-2}
\end{equation}
where $\Delta \epsilon^n$ is the change in the non-wetting phase volume fraction 
that results from the operation $\Omega_n \oplus \varsigma_\delta$. 
Eq. \ref{eq:steiner-2} is a kinematic statement that holds locally; the volume of an object 
only changes as a consequence of net movement of the object boundary. The associated change in volume can be expressed in terms of the invariant properties of the boundary -- surface area, integral mean curvature, and total curvature. 
Based on these arguments, Eq. \ref{eq:steiner-2} provides a geometric relationship to predict
the change in volume for a particular object. 

From the standpoint of model development, it is further necessary to connect geometry to thermodynamics 
so that processes involving energy dissipation can be described. The thermodynamically constrained averaging
theory (TCAT) provides a framework to accomplish this objective. TCAT models are developed by applying rigorous
averaging procedures to directly connect microscale thermodynamics and continuum mechanical forms with their macroscale 
counterparts \cite{Gray_Miller_07,Gray_Miller_13,Gray_Miller_14}. Within TCAT, capillary pressure is the product of the average mean curvature of the interface between fluids, $J_w^{wn}$, and the interfacial tension between the two fluids. Relevant geometric measures are listed in Table \ref{tab:geom}.
Based on these measures, a clear connection to thermodynamics is established:
phase volume fractions and specific interfacial areas are needed to predict the internal energy of the macroscopic system using appropriate thermodynamic expressions. The possible values for these quantities are constrained by geometric laws.

\begin{table}[htbp]
   \centering
   \caption{Relevant geometric measures included in TCAT}
   \begin{tabular}{ll} % Column formatting, @{} suppresses leading/trailing space
      \hline
      Quantity & Description \\
      \hline
      $\epsilon$& Porosity \\
      $\epsilon^{{n}}$ & Volume fraction of the non-wetting fluid \\       	                       
      $\epsilon^{{wn}}$ & Surface area per unit volume for the $wn$ interface  \\
      $\epsilon^{{ns}}$ & Surface area per unit volume for the $ns$ interface \\
      $J_w^{wn}$ & Average mean curvature of $wn$ interface \\
      $J_s^{ns}$ & Average mean curvature of $ns$ interface \\
    \hline
    \end{tabular}
   \label{tab:geom}
\end{table}

The first three Minkowski functionals can be expressed directly in terms of the quantities listed in 
Table \ref{tab:geom},
\begin{eqnarray}
M_0^n &=& \epsilon^{{n}} V\;,
\label{eq:M0-tcat} \\
M_1^n &=& (\epsilon^{{wn}} + \epsilon^{{ns}})V\;,
\label{eq:M1-tcat} \\ 
M_2^n &=& (J_w^{wn}\epsilon^{{wn}} + J_s^{ns}\epsilon^{{ns}})V\;,
\label{eq:M2-tcat}
\end{eqnarray}
where $V$ is the volume of $\Omega$. 
Eq. \ref{eq:steiner-2} can be re-expressed in terms of these relationships.
A non-dimensional form is obtained by introducing the Sauter mean diameter
as a reference length scale,
$D = 6 \epsilon/(\epsilon^{{ns}} + \epsilon^{{ws}})$, and defining $\acute \delta \equiv \delta/D$.
In terms of these alternate quantities, Eq. \ref{eq:steiner-2} becomes
\begin{eqnarray}
\Delta \epsilon^n &=&
a_1 (\acute\epsilon^{wn} + \epsilon^{ns} D ) \acute \delta +  \nonumber \\
&& a_2 (\acute{J}_w^{wn} \acute \epsilon^{wn} + J_s^{ns} \epsilon^{ns}D^2)\acute \delta^2 
+ a_3 \acute{\chi}^n \acute \delta^3\;.
\label{eq:steiner-nondim}
\end{eqnarray}
Three non-dimensional quantities have been introduced,
\begin{eqnarray}
\acute \epsilon^{wn} &=&  \epsilon^{wn} D\;,
\label{eq:awn}\\  
\acute J_w^{wn} &=& J_w^{wn} D\;, \mbox{ and}
\label{eq:Jwn}\\
\acute{\chi} ^n &=& \chi^n D^3 / V\;.
\label{eq:Xn}
\end{eqnarray}
Based on Eq. \ref{eq:steiner-nondim}, different objects will yield different coefficients $a_i$ 
according to Minkowski's quermassintegral \cite{Klain_95}. For example, it is not obvious that two different 
objects with identical Minkowski functionals would also have the same coefficients. Thus
it remains to be demonstrated that Eq. \ref{eq:steiner-2} will produce unique relationships between the Minkowski 
functionals that will hold for two-fluid flow. 
In this work this postulate is tested computationally by considering a very large
number of geometric states. We consider the case where the porosity $\epsilon$ is constant and the role of the fluid
volumes is included from the saturation, $s^w=\epsilon^w/\epsilon=(1-\epsilon^n)/\epsilon$.
We further assume that $J_s^{ns}(s^{{w}},\acute\epsilon^{{wn}},\acute\chi^{{n}})$ and 
$\epsilon^{{ns}}(s^{{w}},\acute\epsilon^{{wn}},\acute\chi^{{n}})$.
We therefore pose the geometric state function as $\acute J_w^{wn} (s^{{w}},\acute\epsilon^{{wn}},\acute\chi^{{n}})$.
This form reflects the fact that the solid is immobile and movement of the non-wetting phase boundary
occurs only at the interface between the two fluids. 
Numerical results used in conjunction with experimental data will be used to show that this relationship
is able to characterize the possible non-wetting fluid configurations in porous media. 

\section{Ink Bottle Example}

\begin{figure}[ht]
  \includegraphics[width=1.0\linewidth]{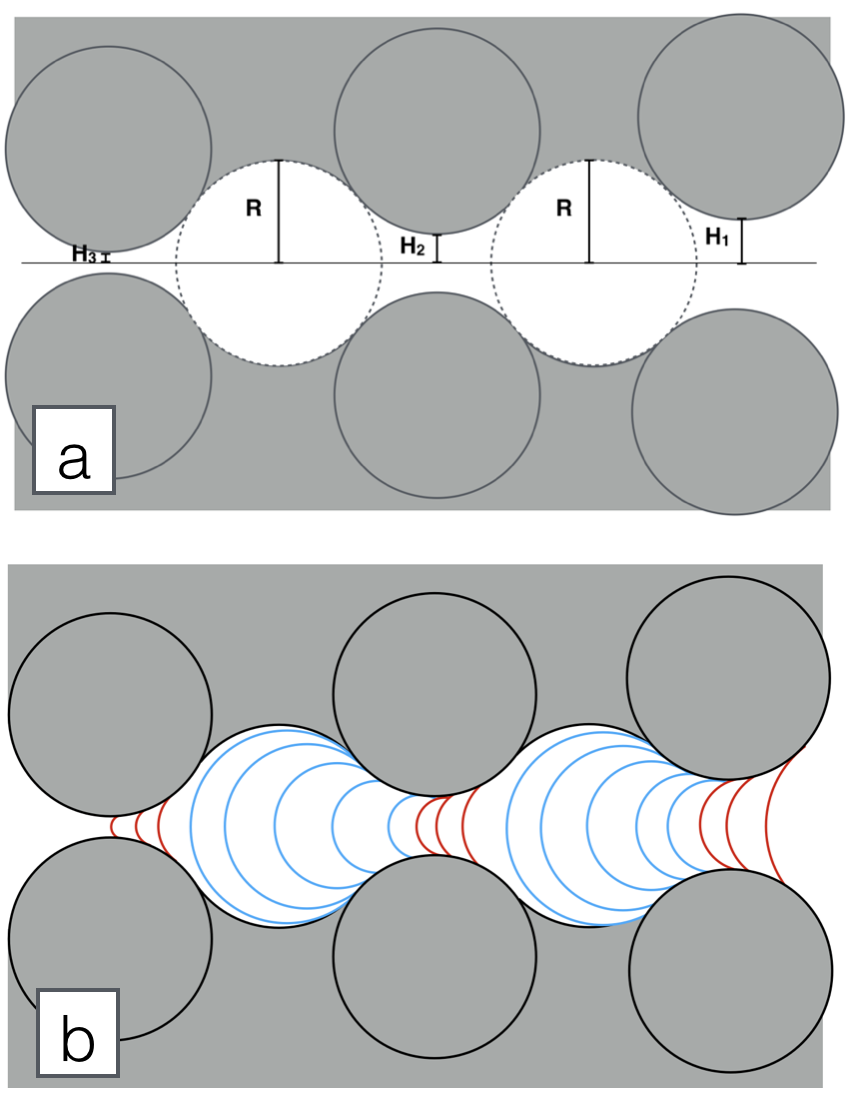}
\caption{
(a) The ink bottle geometry is defined based on the position of 
five equally sized spheres in a symmetric system with radius $R$ to create a two-pore system connected to three throats with
widths $H_1$, $H_2$ and $H_3$. (b) Meniscus configurations within
the ink bottle can be determined analytically. }
\label{fig:inkbottle-Geometry}       % Give a unique label
\end{figure}

Ink bottles such as the example shown in Fig. \ref{fig:inkbottle-Geometry}  are often
used to illustrate the reason for history dependence in the capillary pressure relation.
The ink bottle geometry is a surface-of-revolution of a piecewise smooth
function $\rho(x)>0$, which determines the width of the channel as a function of position $x$. 
The ink bottle shown in Fig. \ref{fig:inkbottle-Geometry} (a) is produced by considering five equally-sized 
circles with radius $R$, positioned to create a sequence of pore bodies connected by 
throats of various sizes. The minimum width for each of the three throats is
$H_1$, $H_2$ and $H_3$, respectively. The $\{x,y\}$ coordinates for the circle centers are given by
\begin{eqnarray}
\bm{c}^{(1)} &=&\{ R, H_1+R \} \\
\bm{c}^{(2)} &=&\{ c^{(1)}_x + \sqrt{4R^2-(H_1+R)^2},0\} \\
\bm{c}^{(3)} &=&\{ c^{(2)}_x + \sqrt{4R^2-(H_2+R)^2},H_2+R\} \\
\bm{c}^{(4)} &=&\{ c^{(3)}_x + \sqrt{4R^2-(H_2+R)^2},0\} \\
\bm{c}^{(5)} &=&\{ c^{(4)}_x + \sqrt{4R^2-(H_3+R)^2},H_3+R\}\;.
\end{eqnarray}
The position of the ink bottle wall is given by $\rho(x)$, which is determined based on the five circles,
\begin{equation}
\rho(x) = \left\{\begin{array}{ll}
c^{(1)}_y  + R \sin \Big(\arccos \frac{x-c^{(1)}_x}{R}\Big) & \mbox{for $x < x_1$} \\
-R \sin \Big(\arccos \frac{x-c^{(2)}_x}{R}\Big) & \mbox{for $x_1 \le x < x_2$} \\
c^{(3)}_y  + R \sin \Big(\arccos \frac{x-c^{(3)}_x}{R}\Big) & \mbox{for $x_2 \le x < x_3$ } \\
-R \sin \Big(\arccos \frac{x-c^{(4)}_x}{R}\Big) & \mbox{for $x_3 \le x < x_4$} \\
c^{(5)}_y  + R \sin \Big(\arccos \frac{x-c^{(5)}_x}{R}\Big) & \mbox{for $x_4 \le x$}  \\
\end{array} \right.,
\end{equation}
where the contact points between the circles are
\begin{eqnarray}
{x}^{(1)} &=& c^{(1)}_x + \sqrt{R^2- (H_1+R)^2/2} \\
{x}^{(2)} &=& c^{(2)}_x + \sqrt{R^2- (H_2+R)^2/2} \\
{x}^{(3)} &=& c^{(3)}_x + \sqrt{R^2- (H_2+R)^2/2} \\
{x}^{(4)} &=& c^{(4)}_x + \sqrt{R^2- (H_3+R)^2/2}\;.
\end{eqnarray}

Simple fluid displacement processes within the ink bottle are often used to illustrate the
origin of geometric hysteresis in porous media.
From the standpoint of an experiment, there are two basic ways to impact the geometric state within a porous 
medium. First, one can can inject fluid into the system from the boundary, explicitly controlling the 
volume of
fluid injected and allowing the fluid pressures to adjust. Second, one can control the pressure for each 
fluid at the boundary,
allowing the fluid volumes to adjust until a stable equilibrium is achieved. Any sensible description 
of the system must be able to describe either scenario. 
We consider the quasi-static case for which the difference in fluid pressures is 
exactly balanced by the capillary pressure based on the Laplace equation.
The fluid pressures are constant within their respective phase regions and the curvature is constant
over the meniscus (i.e. $p^c = -\gamma^{wn} J_w^{wn}$). The macroscale fluid-fluid interfacial tension
is constant and the contact angle is zero degrees.

\begin{figure}[ht]
  \includegraphics[width=1.0\linewidth]{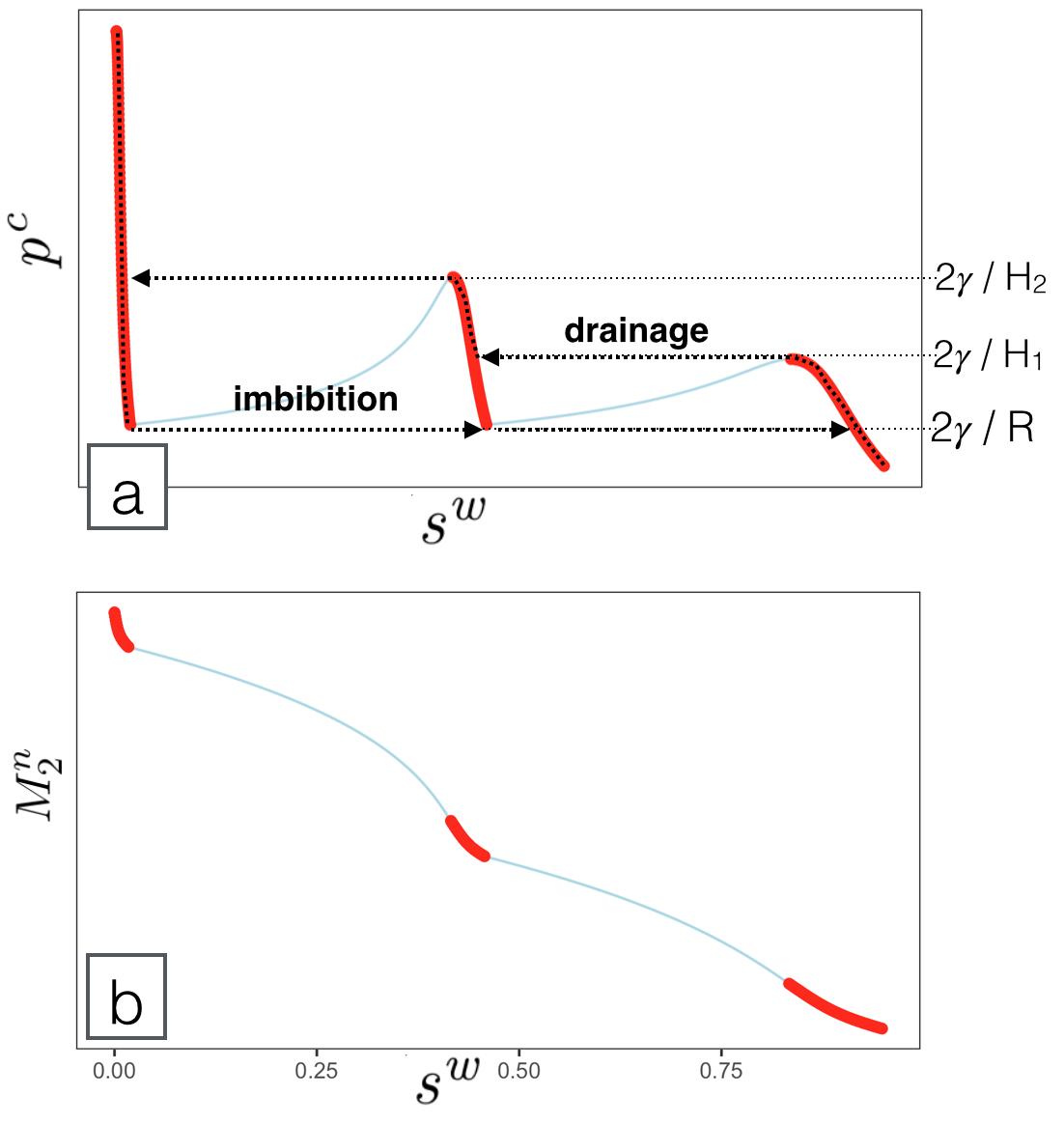}
\caption{Apparent hysteresis in the ink bottle results from
the observation apparently different equilibrium states 
for drainage and imbibition; (b) the integral mean curvature
is a one-to-one function of the saturation. 
}
\label{fig:inkbottle-plots}       % Give a unique label
\end{figure}

Fluid displacement within the ink bottle proceeds with the non-wetting fluid invading from the
right boundary of Fig. \ref{fig:inkbottle-Geometry} as the difference between the fluid pressure increases, 
and the wetting fluid exiting at the left. Geometric measures for the non-wetting fluid are determined based on
the position of the contact line in the $x$ direction.
Meniscus configurations can be determined based on $\rho(x)$ and it's derivative.
When the common curve is at $x$, the meniscus curvature is
\begin{equation}
r(x) = \rho \sqrt{1+\Big(\frac{d\rho}{dx}\Big)^2}\;.
\end{equation}
The height of the spherical cap is 
\begin{equation}
h(x) = r - \rho \Big(\frac{d\rho}{dx}\Big)\;.
\end{equation}
Note that for the quasi-static displacements shown in Fig. \ref{fig:inkbottle-Geometry} (b)
the position of the common curve will always be 
within a pore throat. The non-wetting fluid volume fraction, the interfacial areas per unit volume, 
and mean curvature for each interface are each determined by the position of the common curve, 
\begin{eqnarray}
\epsilon^{n}(x)  &=& \frac{\pi}{V} \Bigg[ \int_0^{x} \rho^2 dx^\prime  + \frac{1}{6} h (3 \rho^2 + h^2) \Bigg] 
\label{eq:inkbottle-1}\\
\epsilon^{wn}(x)  &=& \frac{2 \pi r h}{V} 
\label{eq:inkbottle-2}\\
J_w^{wn}(x)  &=& \frac{2}{r} 
\label{eq:inkbottle-3}\\
\epsilon^{ns}(x) &=& \frac{2 \pi}{V} \int_0^{x} \rho \sqrt{1+\Big(\frac{d\rho}{dx^{\prime}}\Big)^2} dx^\prime 
\label{eq:inkbottle-3}\\
J_s^{ns}(x) &=&  \frac{2 \pi}{\epsilon^{ns} V} \int_0^{x} {\rho} \Bigg(\frac{1}{r_1} + \frac{1}{r_2}\Bigg) \sqrt{1+\Big(\frac{d\rho}{dx^{\prime}}\Big)^2} dx^\prime, 
\label{eq:inkbottle-4}
\end{eqnarray}
where $r_1(x)$ and $r_2(x)$ are the two principal radii of curvature for the ink bottle wall. Since the geometry is a surface of revolution, $r_1=\rho$. Within the pores $r_2=R$. Within the throats $r_2 = -R$.  
Quasi-static displacements within the ink bottle can be described analytically based on these expressions.
As $p_c$ increases, the meniscus squeezes through the pore throats leading to a decrease in $s^w$. 
Pore-filling occurs spontaneously during drainage based on the sequence of 
blue menisci in Fig. \ref{fig:inkbottle-Geometry} (b). The red and blue menisci are associated with the red 
and blue portions of the curve in Fig. \ref{fig:inkbottle-plots} (a) and (b), respectively.
Based on Fig. \ref{fig:inkbottle-plots} (a) the equilibrium meniscus curvature is apparently 
lower during imbibition as compared to drainage. However, since the fluid volume can only change due to movement of the meniscus, changes in the geometric state are kinematic in nature.
While traditionally $p^c(s^w)$ considers only stable equilibrium menisci (red),
under quasi-static conditions the same configurations are obtained independent from flow direction, which includes blue portions of the curve in Fig. \ref{fig:inkbottle-plots} (a) and (b). 

To see how the Minkowski functionals remove the apparent hysteresis in the ink bottle, we consider the
integral mean curvature
\begin{equation}
M_2^n  = (J_w^{wn}\epsilon^{{wn}} + J_s^{ns}\epsilon^{{ns}})V.
\label{eq:M2}
\end{equation}
Fig. \ref{fig:inkbottle-plots} (b) shows that there is a one-to-one relationship between $M_2^n$
and $s^w$. While the curvature $J^{wn}_w$ is not a one-to-one function of $s^w$,
it can be clearly inferred based on $M_2^n$, $\epsilon^{wn}$, $\epsilon^{ns}$, and the curvature of the
solid $J_s^{ns}$.  Since all quantities in Eqs. \ref{eq:inkbottle-1} -- \ref{eq:inkbottle-4} are 
functions of $x$, only a single degree of freedom exists for the ink bottle problem; 
each quantity is uniquely determined by the position of the common curve.
It is also important to note that for the ink bottle geometry the integral mean curvature and surface area 
are not independent quantities since the ink bottle is a surface-of-revolution. Ink bottles are generated by rotating
two-dimensional objects with fewer geometric degrees of freedom; radial symmetry leads to menisci
that are perfect spherical caps. For a spherical meniscus, the surface area is directly determined by the 
curvature. This contrasts with real porous media where the surface area and mean curvature are  
independent quantities.  Furthermore, the ink bottle provides only a single channel for flow
such that complex fluid connectivity is impossible (i.e. $\chi^n = 1$).
This is an important reason why fewer geometric invariants are needed to characterize states within
the ink bottle. It is therefore essential to consider more complex geometries to understand the possible
geometric states within porous media.

\section{Porous media and flow conditions}

%PARAGRAPH 4
% Highlight uCT as emerging data source that can be used to resolve issues with microscale characterization
The microscopic details of fluid and solid materials can be observed experimentally from
$\mu$CT \cite{Berg_Ott_etal_13,Spanne_94,Georgiadis_Berg_etal_13,Armstrong_McClure_etal_16}. 
The resulting data can be used to measure interfacial curvatures, surface areas, the Euler characteristic, and phase connectivity. 
Six different solid materials and relevant properties are listed in Table \ref{tab:media}. These include one synthetic 
sphere pack and five different $\mu$CT images: robuglass, sandpack, 
carbonate rock (Estaillades) and two sandstones (Gildehauser and Castlegate). 
The robuglass, Gildehauser and Estaillades datasets were proccessed using gradient-based segmentation, which is known to provide a valid representation of the resolved rock geometry \cite{schluter2014image}. The 
remaining three data sets are associated with previously published work and are 
publicly accessible \cite{McClure_16,Sheppard_15}. 

For each geometry, we rely on direct numerical simulation within the observed rock geometries to generate a large number of states
as needed to evaluate $\acute J_w^{wn} (s^{{w}},\acute\epsilon^{{wn}},\acute\chi^{{n}})$\cite{McClure_Prins_etal_14}. On the basis of pore-level occupancy and connectivity, close agreement was previously demonstrated between simulated geometries and those observed from experiment using fast $\mu$CT \cite{Armstrong_McClure_etal_16}. Since different physical processes produce different fluid arrangements, multiple simulation conditions were considered, including:
(A) equilibrium simulations with a
random initial condition \cite{McClure_Berrill_etal_16b}; 
(B) steady-state flow with an initial condition from $\mu$CT experiment \cite{Armstrong_McClure_etal_16};
(C) steady-state flow with an initial condition from morphological opening
\cite{Hilpert_01,Adalsteinsson_Hilpert_06,Liu_Herring_etal_17,berg2016connected}; and 
(D) displacement with changing saturation driven by pressure boundary conditions.
The initial conditions and simulation types are listed in Table \ref{tab:media}, along with the total number of distinct fluid configurations generated. Additional descriptions of the simulations are provided
in the Supporting Materials along with the geometric data generated for each material.

\begin{table}[htbp]
   \centering
   \caption{Summary of rock geometries and simulation conditions considered in this work.}
   \begin{tabular}{lccccc} % Column formatting, @{} suppresses leading/trailing space
      \hline
      Media   & $\epsilon$ &$D$(mm) & Size (voxels) & Sim. & Config. \\
      \hline
Castlegate  &  $0.205$ & $0.111$  &$512\times 512 \times 512$  & A,D & 23,123\\
Estaillades &  $0.111 $ & $0.124$ & $834\times 834 \times 556$ & A,B,D & 23,599\\
Gildehauser  & $0.188 $ & $0.133$ &$852\times 852 \times 569$& A,B,D & 38,788 \\
Robuglass & $0.345$ & $0.173$  &$988\times 988\times 598$ & A,B,D & 49,515 \\
Sand pack &  $0.376$ & $0.368$ &$512\times 512 \times 512$ & A,C,D & 64,650 \\
Sphere pack  & $0.369$ & $1.00$  &$900\times 900 \times 900$ & A,C,D & 59,341 \\
      \hline
    \end{tabular}
   \label{tab:media}
\end{table}

%\section{\label{sec:disc}Discussion}
\section{Macroscale Closure Relations}

\begin{figure}[ht]
  \includegraphics[width=1.0\linewidth]{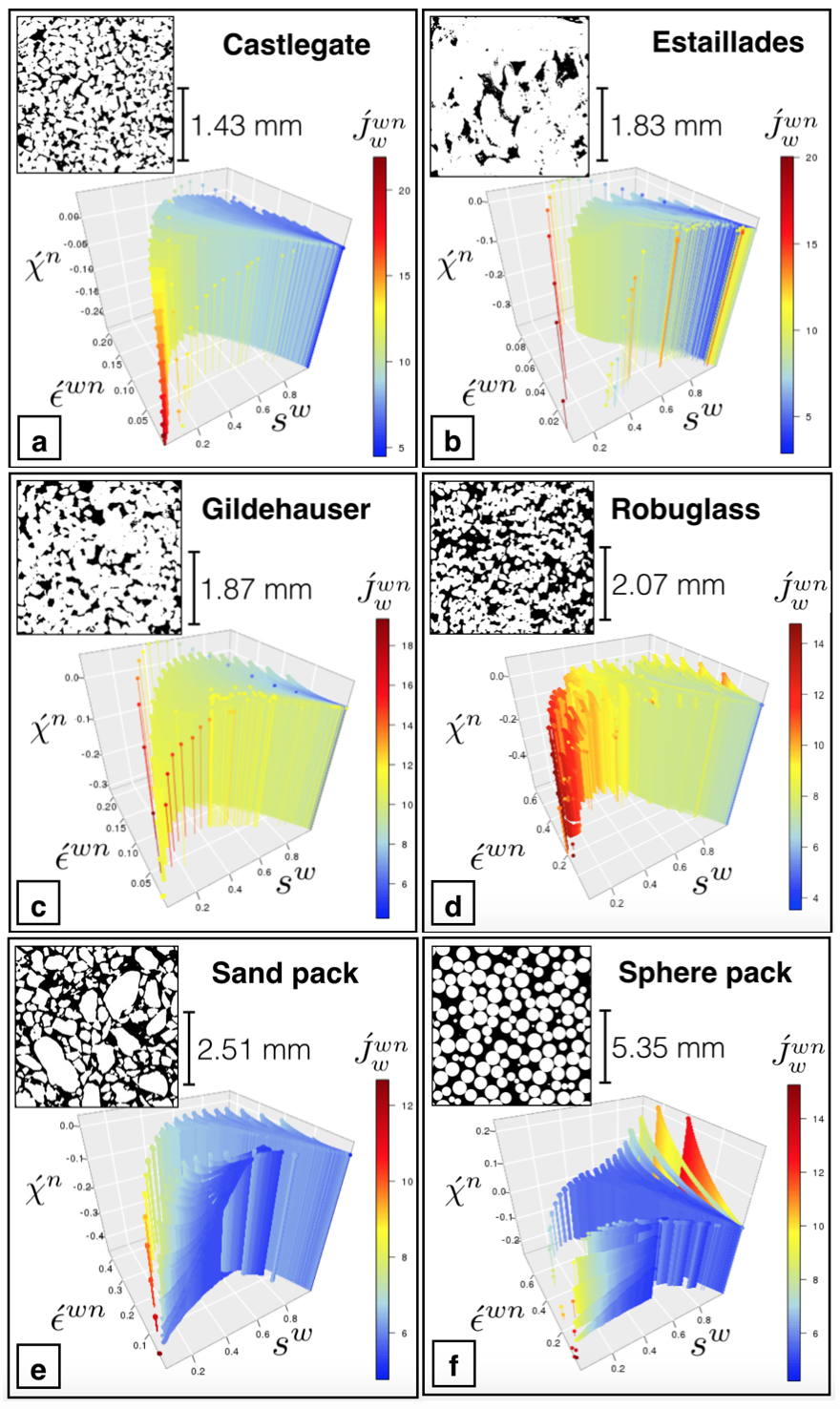}
\caption{The relationship between geometric state variables is  explored based on the 
possible fluid microstates in six porous media, each 
with a distinct solid microstructure (white).
}
\label{fig:psax}       % Give a unique label
\end{figure}

For each porous medium three constitutive relationships were evaluated:
$\acute J_w^{wn}(s^{{w}})$, 
$\acute J_w^{wn}(s^{{w}},\acute\epsilon^{{wn}})$ and 
$\acute J_w^{wn}(s^{{w}},\acute\epsilon^{{wn}},\acute\chi^{{n}})$. 
To adequately assess the uniqueness of a relationship with three possible degrees of freedom, 
a large number of states must be considered. 
%Ideally,multiple measurements of $\acute J_w^{wn}$ should be obtained for 
%each possible set of $s^{{w}},\acute\epsilon^{{wn}}$, and $\acute\chi^{{n}}$.
The present study includes a total of 259,016 fluid configurations for six
different porous media. Averaged measures were determined from simulated 
microstates using an {\em in situ} analysis framework \cite{McClure_Berrill_etal_16a,McClure_Berrill_etal_16c}. The measured values of $\acute J_w^{wn},s^{{w}},\acute\epsilon^{{wn}}$, and $\acute\chi^{{n}}$ are shown in Fig. \ref{fig:psax} (a)--(f). 
Generalized additive models (GAMs) were used to construct locally-smooth spline surfaces to approximate each of the three possible state functions based on the data shown in Fig. \ref{fig:psax}.
The GAMs were used to approximate each relationship, evaluate the error, and to make comparisons among the three approximations \cite{Wood_Goude_etal_15,McClure_Berrill_etal_16b}. 
The GAM can be considered as a numerical approach to evaluate the coefficients in Eq. \ref{eq:steiner} and determine if the relationship is unique for all microstates.

\begin{figure*}[ht]
\centering
\includegraphics[width=1.0\linewidth]{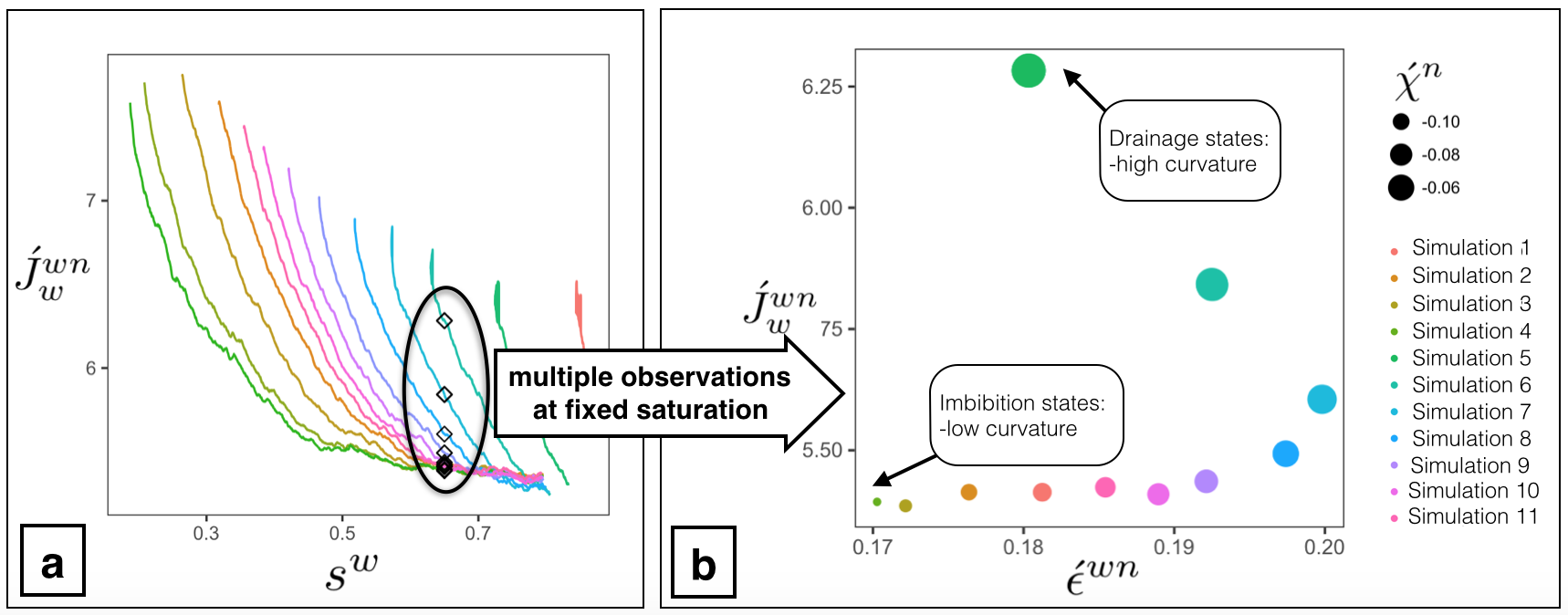}
\caption{Traditional models assume that the macroscale capillary pressure is a function
only of the saturation of the wetting fluid. Two-fluid displacement simulations within 
a sand pack show that:
(a) At fixed saturation, $s^{{w}}$, the relative mean curvature $\acute J_w^{wn}$ can attain 
many possible values depending on the system history;
(b) $\acute J_w^{wn} (s^{{w}},\acute\epsilon^{{wn}})$ is non-unique for $s^{{w}}=0.65$. }
 \label{fig:hysteresis}
\end{figure*}

The non-uniqueness of $\acute J_w^{wn}(s^{{w}})$ and $\acute J_w^{wn}(s^{{w}},\acute\epsilon^{{wn}})$
can be assessed visually based on Fig. \ref{fig:hysteresis}. A collection
of curves are generated by simulating imbibition based on a set of initial conditions determined based on morphological
drainage. In each case, the meniscus curvature decreases during the imbition process, but distinctly different 
geometric states are obtained along each trajectory. The relationship $\acute J_w^{wn}(s^{{w}})$ is clearly non-unique
since multiple values of $\acute J_w^{wn}$ are observed at constant $s^{{w}}$. Furthermore, from  
Fig. \ref{fig:hysteresis} (b) we see that $\acute J_w^{wn}(s^{{w}},\acute\epsilon^{{wn}})$ is also
non-unique; for a set of states with $s^w=0.65$ (one for each curve in Fig. \ref{fig:hysteresis} (a))
we see that the different $\acute J_w^{wn}$ may be observed for the same values of both $s^w$ and 
$\acute\epsilon^{{wn}}$. Examining the Euler characteristic demonstrates that different fluid connectivity is 
observed within this set of states, suggesting the underlying reason for the behavior.
This also illustrates the importance of generating redundant geometric states to test the uniqueness of 
the geometric relationship. A limiting factor in previous studies is that an insufficient number of points
were generated for a particular volume fraction. 

\begin{figure}[ht]
  \includegraphics[width=1.0\linewidth]{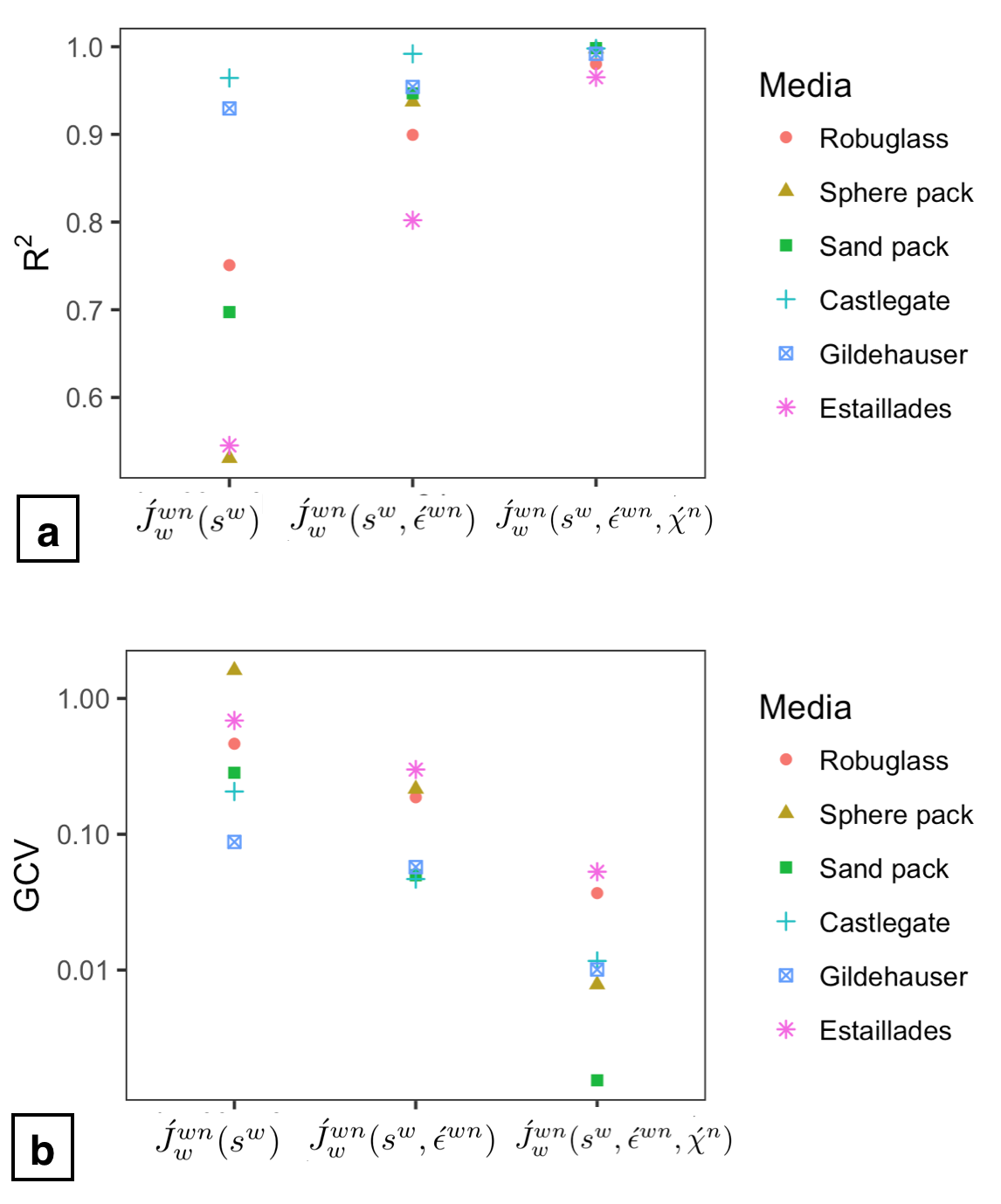}
\caption{The geometric state function $\acute J_w^{wn}(s^{{w}},\acute\epsilon^{{wn}},\acute\chi^{{n}})$ is unique for each geometry, which is not the case for $\acute J_w^{wn}(s^{{w}})$
and $\acute J_w^{wn}(s^{{w}},\acute\epsilon^{{wn}})$. 
Locally-smooth approximations were generated for each relationship using generalized additive models (GAMs) to compare the accuracy of macroscopic constitutive models.
Error measures are given by (a) the coefficient of variation $R^2$; and 
(b) generalized cross validation (GCV).}
\label{fig:model}       % Give a unique label
\end{figure}

In Fig. \ref{fig:model} two different error measures are used to assess each model.
Generalized cross-validation (GCV) 
considers how accurately each data point is predicted by all remaining data points \cite{Golub_79}. 
The GCV will be $0.0$ if the surface perfectly predicts the data. A second measure is the coefficient of determination,
$\mbox{R}^2$, which measures the fraction of the variance in the underlying data points predicted
based on the GAM. The value of $\mbox{R}^2$ will be $1.0$ for a unique surface. The fact that $\acute J_w^{wn}(s^{{w}})$ is not-unique for any of the six materials is confirmed by the large GCV ($0.0878 \le \mbox{GCV} \le 1.613$) and a relatively small $\mbox{R}^2$ ($  0.53 \le \mbox{R}^2 \le 0.96$). The sphere pack, sand pack and carbonate allow for the widest range of possible fluid configurations at any given saturation, and show the largest unexplained variance for $\acute J_w^{wn}(s^{{w}})$.
The two sandstones and robuglass samples show the least variance.
Traditionally the unexplained variance would be attributed to hysteresis. In all cases, the unexplained variance is reduced by including the interfacial area, $\acute J_w^{wn}(s^{{w}},\acute\epsilon^{{wn}})$, with  $0.057 \le \mbox{GCV} \le 0.30$ and $ 0.80 \le \mbox{R}^2 \le 0.99$. When all three independent MFs are included, $\acute J_w^{wn}(s^{{w}},\acute\epsilon^{{wn}},\acute\chi^{{n}})$, the accuracy improves dramatically with
$0.0015 \le \mbox{GCV} \le 0.053$ and $ 0.96 \le \mbox{R}^2 \le 1.0$. 
The only material for which $\acute J_w^{wn}(s^{{w}},\acute\epsilon^{{wn}},\acute\chi^{{n}})$ captures less than $99$\% of the variance in the underlying data is the carbonate, shown in Fig. \ref{fig:psax} (b).
Since $\acute J_w^{wn}(s^{{w}},\acute\epsilon^{{wn}})$ explains only 80\% of the variance for fluid configurations within the carbonate, this is also the material for which adding the Euler characteristic provides the most significant improvement. The implication is that a significant fraction of the variance 
in the carbonate is due to changes in fluid connectivity that are not predicted by changes in interfacial area.

%\begin{figure*}[htbp]

In all six cases, $\acute J_w^{wn}(s^{{w}})$ is observed to be non-unique. 
While this is mitigated when using $\acute J_w^{wn}(s^{{w}},\acute\epsilon^{{wn}})$, the Euler characteristic must be included to obtain a unique relationship. Non-uniqueness in $\acute J_w^{wn}(s^{{w}},\acute\epsilon^{{wn}},\acute\chi^{{n}})$ is observed only when $s^{{w}}\rightarrow 1$. This occurs because the quantities $\acute\epsilon^{{wn}},\acute\chi^{{n}}$ approach zero in the limit of vanishing saturation, whereas no limit exists for $\acute J_w^{wn}$  \cite{McClure_Berrill_etal_16b}. Numerical errors associated with the measurement of the interfacial area, curvature, and Euler characteristic are expected to account for a contribution to the error. Since the experimental and simulated non-wetting fluid configuration make use of a closed three-dimensional object, errors due to system size do not undermine the effort to test the validity of $\acute J_w^{wn}(s^{{w}},\acute\epsilon^{{wn}},\acute\chi^{{n}})$. 

%\begin{figure*}[htbp]

The notion of a representative elementary volume (REV) 
is important to produce generalizable geometric state information based on microscopic measurements. 
In this work, we have shown that the form
$\acute J_w^{wn}(s^{{w}},\acute\epsilon^{{wn}},\acute\chi^{{n}})$
is valid based on non-wetting fluid configurations in water-wet porous media
based on state-of-the-art $\mu$CT data. While the physical size of the
systems considered in this work remains relatively small, future 
advancement of $\mu$CT and computing technologies will lead to growth in 3D image sizes that will continue for the foreseeable future. 
Based on on this trajectory, REV-scale observations of porous medium micro-structure are likely. 
These trends favor the development new models that are able to leverage the additional geometric information that is now available for flow in porous media. While factors such as length-scale heterogeneity will remain an important consideration for flow in porous media (i.e. characterization of
geologic reservoirs), macroscopic strategies to address this problem are already under active development.

\section{Conclusions}

We present a geometric state function to predict the mean curvature of the interface between fluids in porous medium based on a relationship between invariants established by integral geometry. The relationship accurately captures a broad range of possible fluid configurations, and can be combined with modern averaging methods to formulate a new class of model that evolves interfacial areas using conservation equations and kinematic equations, applies to all water saturation levels, and does not require equilibrium assumptions \cite{Gray_Miller_13,Gray_Miller_14,Gray_Dye_etal_15}. To take full advantage of the equation of state developed, an evolution equation for the Euler characteristic would be needed. It is expected that this development and the proven need for this remaining quantity will catalyze efforts to develop the missing kinematic relation for the Euler characteristic. The geometric result suggests that it is possible to overcome
shortcomings associated with empirical and history-dependent constitutive relationships for
two-fluid flow in porous media.

\begin{acknowledgments}
`This work was supported by Army Research Office grant W911NF-14-1-02877 and National Science Foundation grant 1619767. We acknowledge the Paul Scherrer Institut, Villigen, Switzerland for provision of synchrotron radiation beamtime at beamline TOMCAT of the SLS and would like to thank Kevin Mader and Federica Marone for assistance, and Shell for giving access to the data and supporting publication of this work. An award of computer time was provided by the Department of Energy INCITE program. This research also used resources of the Oak Ridge Leadership Computing Facility, which is a DOE Office of Science User Facility supported under Contract DE-AC05-00OR22725. 
\end{acknowledgments}

\end{document}